\documentclass[12pt, a4paper]{article}
\usepackage{amssymb, amsmath, bm, float, epsfig, color, slashed, subcaption, multirow, arydshln, mathtools, bbm}
\usepackage[centering]{geometry}
\makeatletter
\@addtoreset{equation}{section}
\makeatother

\renewcommand{\Re}{\mathrm{Re}\,}
\renewcommand{\Im}{\mathrm{Im}\,}
\renewcommand{\|}{\hspace{1pt}|\hspace{1pt}}

\begin{document}

\thispagestyle{empty}

\begin{flushright}
April 2021 \\
\end{flushright}
\vspace{30pt}
\begin{center}
{\Large\bf Modular symmetry and zeros\\[10pt] in magnetic compactifications} \\

\vspace{47.5pt}
{\large Yoshiyuki Tatsuta} \\

\vspace{12.5pt}
{\it Scuola Normale Superiore and INFN, Piazza dei Cavalieri 7, 56126 Pisa, Italy}
\end{center}

\vspace{25pt}
\begin{abstract}
\noindent
We discuss the modular symmetry and zeros of zero-mode wave functions on two-dimensional torus $T^2$ and toroidal orbifolds $T^2/\mathbb{Z}_N$ ($N=2,3,4,6$) with a background homogeneous magnetic field.
As is well-known, magnetic flux contributes to the index in the Atiyah-Singer index theorem.
The zeros in magnetic compactifications therefore play an important role, as investigated in a series of recent papers.
Focusing on the zeros and their positions, we study what type of boundary conditions must be satisfied by the zero modes after the modular transformation.
The consideration in this paper justifies that the boundary conditions are common before and after the modular transformation.
\end{abstract}

\newpage
\setcounter{page}{2}
\setcounter{footnote}{0}


\section{Introduction}
Orbifold compactifications play an important role in string theory and higher-dimensional field theory \cite{Dixon:1985jw, Dixon:1986jc}.
This is because they lead to a chiral spectrum in their four-dimensional (4d) effective theories.
The same can be achieved by magnetic flux compacitifications \cite{Abouelsaood:1986gd, Bachas:1995ik}.
In addition to the chiral spectrum, the ground states are degenerate.
The degeneracy can therefore be identified as a family replication like the quarks and leptons in the Standard Model of particle physics.
Based on such a feature, a number of phenomenological applications have been explored in the context of type-I and II string theory \cite{Blumenhagen:2000wh, Angelantonj:2000hi} as well as the field theory \cite{Cremades:2004wa, Abe:2013bca},\footnote{See for review \cite{Angelantonj:2002ct, Blumenhagen:2005mu, Blumenhagen:2006ci, Ibanez:2012zz}.} e.g., three-generation models \cite{Abe:2008sx, Abe:2015yva}, flavor hierarchy and mixings \cite{Abe:2014vza, Fujimoto:2016zjs, Kobayashi:2016qag, Buchmuller:2017vho, Buchmuller:2017vut}, discrete flavor symmetry \cite{Abe:2009vi, BerasaluceGonzalez:2011wy, Marchesano:2013ega, Abe:2014nla} and other phenomenological aspects \cite{Buchmuller:2016gib, Ghilencea:2017jmh, Buchmuller:2018eog, Lim:2018lgg, Hirose:2019ywp, Buchmuller:2019ipg}.

A well-known fact is that a background magnetic field contributes to the index in the Atiyah-Singer index theorem \cite{Atiyah:1963zz}.
For two-dimensional (2d) torus, the number of flux quanta $M$ is equal to the index \cite{Green:1987mn, Witten:1984dg}
\begin{align}
{\rm Ind}\,(i \slashed{D}) &= n_{+} - n_{-} \notag \\
&= \frac{q}{2\pi} \int_{T^2} F = M.
\end{align}
This theorem can be expressed in terms of winding numbers around zeros \cite{VENUGOPALKRISHNA1972349, Weinberg:1981eu}
\begin{gather}
{\rm Ind}\,(i \slashed{D}) = \sum_{i} \frac1{2\pi i} \oint_{C_i} \bm{\nabla} (\log \, \xi^j(z, \tau)) \cdot d \bm{\ell},
\end{gather}
where $i$ labels zeros of the ground-state mode functions $\xi^j(z, \tau)$.
The zeros in the system of magnetic compactifications therefore play a crucial role.
In particular, on the orbifolds the particular zeros are associated with orbifold fixed points where flux is localized \cite{Buchmuller:2015eya, Buchmuller:2018lkz}.
Inspired by the observation, a single zero-mode counting formula on the orbifolds has been found in the previous paper \cite{Sakamoto:2020pev}, which indicates the number of independent orbifold zero modes for any pattern.\footnote{See \cite{Sakamoto:2021nsj} in the case without a background flux.}

A reasonable expectation in various aspects is that a background homogeneous magnetic field may keep the modular symmetry, which has partially been confirmed in \cite{Kikuchi:2020frp}.
As for phenomenological applications in the magnetic compactifications, {\em modular flavor symmetries} have gathered much attention recently.
In particular, it is interesting to investigate what flavor symmetry is derived from ultraviolet setups to their low-energy spectra \cite{Kobayashi:2018rad, Ohki:2020bpo, Kikuchi:2020frp, Hoshiya:2021nux}.
By contrast, infrared (bottom-up) approaches seek to reproduce the observed (lepton) mixings and predict unseen observables \cite{Feruglio:2017spp, Kobayashi:2018scp, Penedo:2018nmg, Novichkov:2018nkm, Ding:2019xna, Liu:2019khw, Chen:2019ewa}.

In a series of papers \cite{Ohki:2020bpo, Kikuchi:2020frp, Almumin:2021fbk}, there has appeared a puzzle concerning boundary conditions: whether the wave functions satisfy the common (pseudo-)periodicity conditions before and after the modular transformation or not.
In this paper, we clarify the puzzle, concentrating on the zeros of the wave functions in the magnetic compactifications.
Several circumstantial evidences suggest that the wave functions satisfy the original periodicity conditions even after the modular transformation.
We then confirm that the suggestion agrees with the zero-mode counting formula mentioned above.
That is one of our main subjects in this paper.

This paper is organized as follows.
In Sec.\,2, we briefly review the zero-mode wave functions on the 2d torus and orbifolds with a background magnetic field.
In Sec.\,3, the Atiyah-Singer index theorem in terms of winding numbers is explained and we classify zero points of the wave functions.
In Sec.\,4, after reviewing the modular group, we look at how the wave functions transform under the modular transformation.
In Sec.\,5, we examine a relation among the original zeros and their zero-point equivalence under two identical transformations.
We find a consequence that the Scherk-Schwarz twists should appropriately shift under the modular transformation.
In Sec.\,6, we mention the zero-mode counting formula on the orbifolds and its modular invariance.
There, it necessitates that the consequence in Sec.\,5 is applied.
Section 7 is devoted to conclusion.
We summarize our notation of the gamma matrices in Appx.\,A and the properties of Jacobi theta-functions in Appx.\,B which are used in the main sections.
In Appx.\,C, the explicit expressions for \eqref{6.6} are listed thoroughly.

\section{Zero modes on magnetized torus and orbifolds}
We start with briefly reviewing zero modes on 2d torus $T^2$ and orbifolds
$T^{2}/\mathbb{Z}_{N}$ ($N=2,3,4,6$) with a background (homogeneous) magnetic field \cite{Cremades:2004wa, Abe:2013bca}.

\subsection{Compactification and background fields}
We consider a six-dimensional (6d) gauge theory compactified on $T^2$ and $T^2/\mathbb{Z}_N$.
For a complex coordinate $z \equiv y_1 + \tau y_2$  ($\tau \in \mathbb{C}, \Im\tau > 0$), the 2d torus $T^2$ is obtained by an identification under torus translations, i.e., $z \sim z + 1 \sim z + \tau$.
As formulated in \cite{Cremades:2004wa}, the 1-form vector potential
\begin{gather}
A(z) \equiv \frac{f}{2 \, \Im \tau} \Im (\bar z dz)
\end{gather}
provides a background homogeneous magnetic field $F=dA$.
The 1-form defined above is not invariant under lattice translations. 
The difference therefore defines a necessary gauge transformation, $A (z + 1) = A (z) + d \Lambda_1(z)$ and $A (z + \tau) = A (z) + d \Lambda_2(z)$, where gauge parameters $\Lambda_1$ and $\Lambda_2$ are given as
\begin{gather}
\Lambda_1(z) = \frac{f}{2 \, \Im \tau} \Im z, \qquad 
\Lambda_2(z) = \frac{f}{2 \, \Im \tau} \Im (\bar \tau z).
\label{transition}
\end{gather}
As stated in \cite{Abouelsaood:1986gd}, the gauge transformation on $T^2$ is well-defined only when the homogeneous flux $f$ is quantized for a given $U(1)$ charge $q$ as
\begin{gather}
\frac{qf}{2 \pi} \equiv M \in \mathbb{Z}.
\label{quantization}
\end{gather}
On the orbifolds $T^2/\mathbb{Z}_N$ one has to notice the presence of localized flux sources at orbifold fixed points \cite{Buchmuller:2015eya, Buchmuller:2018lkz}.
The localized fluxes contribute to Wilson loops around the fixed points and are not ignorable.
It turned out in \cite{Buchmuller:2018lkz} that the transition functions \eqref{transition} and the quantization condition \eqref{quantization} are well-defined even on $T^2/\mathbb{Z}_N$.

We now turn to a 6d Weyl spinor $\Psi$ and its transformation property under a background flux.
The Lagrangian reads
\begin{gather}
{\cal L}_{\rm 6d} = i \bar \Psi \Gamma^M D_M \Psi,
\end{gather}
where $M = \mu \, (= 0, 1, 2, 3), 5, 6$ labels the 6d spacetime, $\Gamma^M$ denotes the 6d gamma matrices (defined in Appx.\,A) and $D_M = \partial_M - iq A_M$ is the covariant derivative.
We decompose the 6d Weyl spinor $\Psi(x, z)$ into 4d left- and right-handed Weyl spinors $\psi^{(4)}_{L}(x)$ and $\psi^{(4)}_{R}(x)$,
\begin{gather}
\Psi(x, z) = \sum_{n, \hspace{.5pt} j} 
\bigl( \psi^{(4)}_{R, \hspace{.5pt} n, \hspace{.5pt} j}(x) \otimes \psi^{(2)}_{+, \hspace{.5pt} n, \hspace{.5pt} j}(z) + \psi^{(4)}_{L, \hspace{.5pt} n, \hspace{.5pt} j}(x) \otimes \psi^{(2)}_{-, \hspace{.5pt} n, \hspace{.5pt} j}(z) \bigr).
\end{gather}
For convenience, 2d Weyl spinors are chosen as
\begin{gather}
\psi^{(2)}_{+, \hspace{.5pt} n, \hspace{.5pt} j} = 
\begin{pmatrix}
\psi_{+, \hspace{.5pt} n, \hspace{.5pt} j} \\[3pt] 0
\end{pmatrix}
, \qquad \psi^{(2)}_{-, \hspace{.5pt} n, \hspace{.5pt} j} = 
\begin{pmatrix}
0 \\[3pt] \psi_{-, \hspace{.5pt} n, \hspace{.5pt} j}
\end{pmatrix},
\end{gather}
where $n$ and $j$ label the Landau level and the degeneracy in each level, respectively.
Note that there exist multiple states labeled by $j$ in each level due to the presence of a non-trivial background magnetic field.

The gauge transformation induced by lattice translations requires the 2d Weyl spinors to satisfy the (pseudo-)periodicity conditions 
\begin{gather}
\psi_{\pm, \hspace{.5pt} n, \hspace{.5pt} j}(z + 1) = U_1(z) \psi_{\pm, \hspace{.5pt} n, \hspace{.5pt} j}(z), \qquad 
\psi_{\pm, \hspace{.5pt} n, \hspace{.5pt} j}(z + \tau) = U_2(z) \psi_{\pm, \hspace{.5pt} n, \hspace{.5pt} j}(z),
\label{BC_torus}
\end{gather}
where the transition functions including the Scherk-Schwarz (SS) twists $\alpha_1$ and $\alpha_2$ are given by
\begin{gather}
U_i(z) = e^{i q \Lambda_i (z)} e^{2\pi i \alpha_i} \quad (i=1,2).
\end{gather}
In general, the twists ($\alpha_1, \alpha_2$) take arbitrary values, $\alpha_1, \alpha_2 \in [0, 1)$ on $T^2$.
On the other hand, they are restricted to specific values on $T^2/\mathbb{Z}_N$, as explained later.

\subsection{Zero modes on $T^2$}
We show the torus zero-mode solutions \cite{Cremades:2004wa}.
In this paper, we restrict the considerations to $M > 0$.
One can obtain the corresponding solutions for $M < 0$ in a similar way (cf. \cite{Cremades:2004wa}).
We concentrate on the ground states, $n=0$, and drop the subscript for simplicity in the following.

The zero-mode equations read
\begin{gather}
\left(\bar \partial + \frac{\pi M}{2 \, \Im \tau} z \right) \psi_{+, j}(z) = 0, \qquad 
\left(\partial - \frac{\pi M}{2 \, \Im \tau} \bar z \right) \psi_{-, j}(z) = 0
\end{gather}
with the (pseudo-)periodicity conditions \eqref{BC_torus} imposed.
For $M > 0$, there exist $M$-fold normalizable solutions only for $\psi_+$,\footnote{For $M < 0$, one finds $|M|$-fold solutions only for $\psi_-$.}
\begin{align}
\psi_{+, \hspace{.5pt} j}(z) &= \mathcal{N} e^{i \pi M z \, \Im z / \Im \tau} \, \vartheta
\begin{bmatrix}
\tfrac{j + \alpha_1}M \\[3pt] -\alpha_2
\end{bmatrix}
(Mz, M\tau) \notag\\
&\equiv \xi^j(z, \tau \| \alpha_1, \alpha_2),
\label{sol_torus}
\end{align}
where $j = 0, 1, ..., M-1 \in \mathbb{Z}_M$ indicates the degeneracy of the zero-mode solutions,\footnote{We define $\mathbb{Z}_M \equiv \mathbb{Z}/M\mathbb{Z} = \{0, 1, \ldots, M-1\}$.} and $\vartheta$ denotes the Jacobi theta-function defined in Appx.\,B.
The normalization factor $\mathcal{N}$ is given by
\begin{gather}
\mathcal{N} = \left( \frac{2 M \Im \tau}{\mathcal{A}^2} \right)^{1/4}
\end{gather}
with the torus volume $\mathcal{A} = (2 \pi R)^2 \Im \tau$.
It is important that profiles of the above wave functions depend on a complex structure modulus $\tau$, the number of flux quanta $M$, the family label $j$ and the SS twists $\alpha_i$ ($i=1,2$).

One should note the the 6d Weyl spinor contains only $M$ 4d chiral spinors $\psi^{(4)}_{R, \hspace{.5pt} 0, \hspace{.5pt} j}(x)$ for $M > 0$. 
In other words, this is the expected consequence that the low-energy spectrum below a compactification scale $M_C \sim 1/R$ is chiral and degenerate as the quarks and leptons in the Standard Model.


\subsection{Zero modes on $T^2/\mathbb{Z}_N$}
The orbifolds $T^2/\mathbb{Z}_N$ ($N = 2, 3, 4, 6$) are obtained by further identifying $z \sim \omega z \qquad$ ($\omega \equiv e^{2 \pi i/N}$).
We first mention possible values of the complex modulus parameter $\tau$ on the orbofolds.
It is discussed in \cite{Choi:2006qh} that $\tau$ can be arbitrary for $N=2$ as long as $\Im \tau > 0$.
By contrast, a consistency based on crystallography demands $\tau = \omega$ for $N = 3, 4, 6$.
The orbifold fixed points, which are invariant under the orbifold identification and lattice translations, are given by
\begin{align}
(y_{1},y_{2})
 = 
\begin{cases}
(0,0), (1/2, 0), (0, 1/2), (1/2, 1/2) & \quad \textrm{on} ~ T^{2}/\mathbb{Z}_{2},\\ 
(0,0), (2/3, 1/3), (1/3, 2/3) & \quad \textrm{on} ~ T^{2}/\mathbb{Z}_{3},\\ 
(0,0), (1/2, 1/2) & \quad \textrm{on} ~ T^{2}/\mathbb{Z}_{4},\\
(0,0) & \quad \textrm{on} ~ T^{2}/\mathbb{Z}_{6}.
\label{fixedpoint}
\end{cases}
\end{align}

In general, the orbifold wave functions transform under the $\mathbb{Z}_N$ rotation $z \to \omega z$ as
\begin{gather}
\psi_{+, \hspace{.5pt} n, \hspace{.5pt} j}(\omega z) = \eta \, \psi_{+, \hspace{.5pt} n, \hspace{.5pt} j}(z), \qquad \psi_{-, \hspace{.5pt} n, \hspace{.5pt} j}(\omega z) = \omega \eta \, \psi_{-, \hspace{.5pt} n, \hspace{.5pt} j}(z),
\label{BC_orbifold}
\end{gather}
where $\eta \equiv \omega^\ell$ ($\ell=0,1,...,N-1)$ classifies the $\mathbb{Z}_N$ parity of the wave functions.
The gauge transformations $U_1(z)$ and $U_2(z)$ in \eqref{BC_torus} and the $\mathbb{Z}_N$ eigenvalue $\eta$ (or $\omega \eta$) have to make a closed group in the wavefunction space \cite{Abe:2013bca}.
As slightly mentioned before, the possible SS twists are restricted by this consistency as
\begin{gather}
(\alpha_{1}, \alpha_{2})
 = (0,0), (1/2,0), (0,1/2), (1/2,1/2) \qquad \textrm{on} ~ T^{2}/\mathbb{Z}_{2},
\label{Z2_SSphase}  \\
\alpha = \alpha_{1} = \alpha_{2} = 
\begin{cases} 
0, 1/3, 2/3 & \quad (M = \textrm{even})\\
1/6, 3/6, 5/6 & \quad (M = \textrm{odd})
\end{cases} \qquad \textrm{on} ~ T^2/\mathbb{Z}_3,
\label{Z3_SSphase} \\
\alpha = \alpha_{1} = \alpha_{2}
 = 0, 1/2 \qquad \textrm{on} ~ T^{2}/\mathbb{Z}_{4},
\label{Z4_SSphase} \\
\alpha = \alpha_{1} = \alpha_{2} = 
\begin{cases} 
0 & \quad (M = \textrm{even})\\
1/2 & \quad (M = \textrm{odd})
\end{cases} \qquad \mathrm{on} ~ T^2/\mathbb{Z}_6.
\label{Z6_SSphase}
\end{gather}

For $M > 0$, one obtains the orbifold eigenstates satisfying \eqref{BC_torus} and \eqref{BC_orbifold} as
\begin{gather}
\xi^j_{\eta} (z, \tau \| \alpha_1, \alpha_2) = \mathcal{N}^{j}_{\eta} \, \sum_{\ell = 0}^{N-1} \bar \eta^\ell \, \xi^j(\omega^\ell z, \tau \| \alpha_1, \alpha_2),
\label{sol_orbifold}
\end{gather}
with the adjusted normalization factor $\mathcal{N}^j_\eta$.

Before closing this section, we mention the number of independent physical ground states on the orbifolds $T^2/\mathbb{Z}_N$.
On the torus $T^2$ there are $M$ independent zero modes, as seen from \eqref{sol_torus}.
On the orbifolds, some modes are projected out by splitting the toroidal zero modes into the orbifold eigen states.
In general, the number of orbifold eigen modes is less than that on the torus (cf.\,\cite{Abe:2013bca}).

\section{Index theorem and zero points}
\subsection{Index theorem on $T^2$}
\begin{figure}[!t]
\centering
\includegraphics[width=0.8\textwidth]{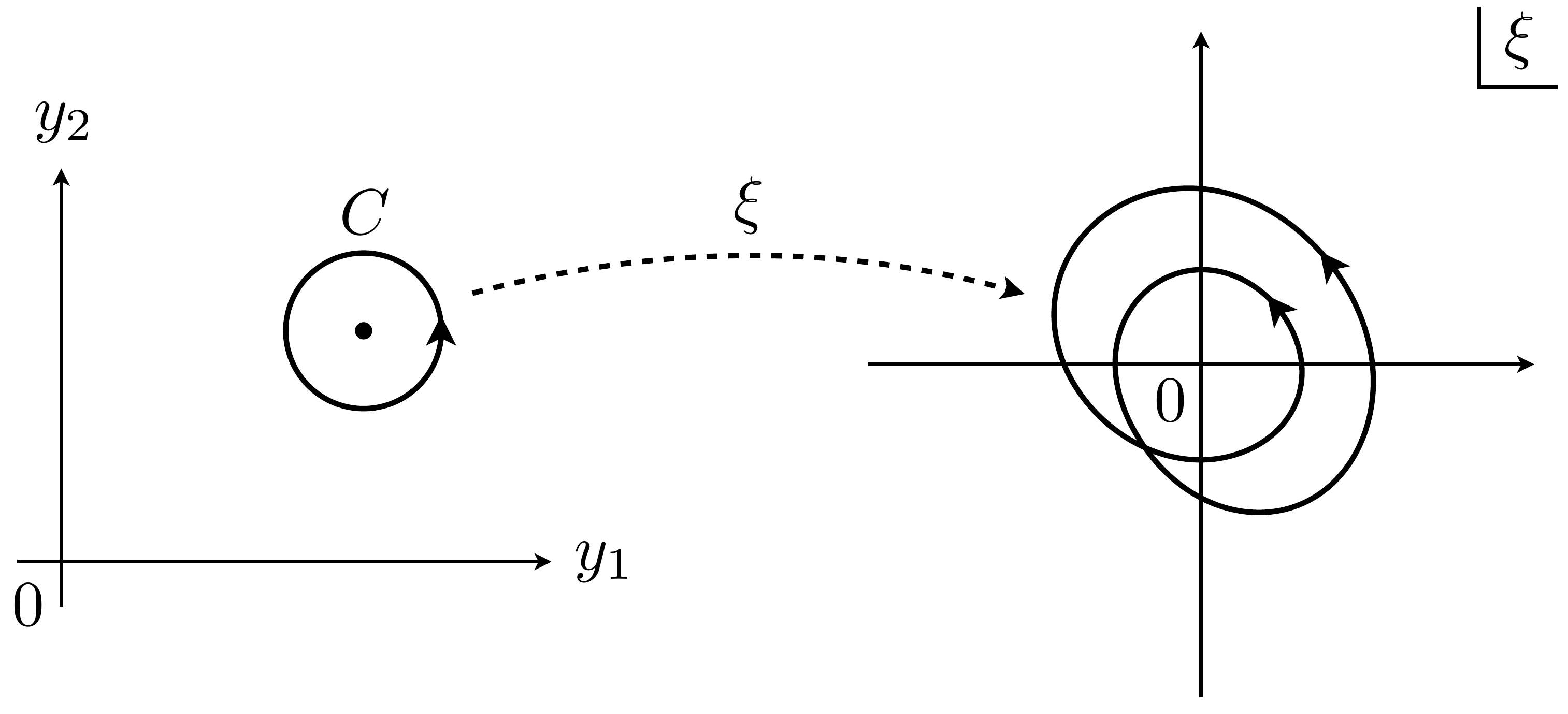}
\caption{Winding number (or vortex number) defined in \eqref{contour_integral}. In this example, the winding number is $+2$. A black dot in the left figure stands for a zero point of $\xi^j(z, \tau)$.}
\label{fredholm}
\end{figure}

It is well-known that the index on $T^2$ with a background flux is identical to the number of flux quanta $M$ \cite{Green:1987mn, Witten:1984dg}
\begin{align}
{\rm Ind}\,(i \slashed{D}) &= n_{+} - n_{-} \notag \\
&= \frac{q}{2\pi} \int_{T^2} F = M,
\label{ASindextheorem}
\end{align}
where $n_\pm$ denotes the number of zero modes for $\psi_\pm$, respectively.
The discussion in Sec.\,2 confirms that the Atiyah-Singer index theorem \eqref{ASindextheorem} holds since $n_+ = M$ and $n_- = 0$ for $M > 0$.\footnote{Similarly, one can show $n_+ = 0$ and $n_- = |M|$ for $M < 0$. See \cite{Cremades:2004wa}.}
As shown in \cite{VENUGOPALKRISHNA1972349, Weinberg:1981eu}, zeros of the zero modes play a crucial role in the index theorem.
Indeed, the index is obtained by the total {\em winding numbers} $\chi_i$ at zeros $p_i$ such that $\xi^j (z = p_i, \tau) = 0$, i.e.,
\begin{align}
{\rm Ind}\,(i \slashed{D}) &= \sum_{i} \frac1{2\pi i} \oint_{C_i} \bm{\nabla} (\log \, \xi^j(z, \tau)) \cdot d \bm{\ell} \notag\\
&\equiv \sum_i \chi_i.
\label{contour_integral}
\end{align}
Here $C_i$ denotes the anti-clockwise contour around the zeros $p_i$ chosen in the fundamental domain of $T^{2}$.
The winding number $\chi_i$ counts how many times $\xi^j$ wraps around the origin, as illustrated in Fig.\,\ref{fredholm}.

\subsection{Zeros on $T^2$}
\label{zero_torus}
First, it is instructive to look at zeros of the torus zero modes \eqref{sol_torus},
\begin{gather}
\xi^j(z, \tau \| \alpha_1, \alpha_2) = \mathcal{N}\,e^{i \pi M z \, \Im z / \Im \tau} \, \vartheta
\begin{bmatrix}
\tfrac{j + \alpha_1}M \\[3pt] -\alpha_2
\end{bmatrix}
(Mz, M\tau).
\end{gather}
Since the Gaussian(-like) exponential factor is always non-zero, zeros originate from the Jacobi theta-function.
Namely, solving an equation
\begin{gather}
\vartheta
\begin{bmatrix}
\tfrac{j + \alpha_1}M \\[3pt] -\alpha_2
\end{bmatrix}
(Mz, M\tau) \stackrel{!}{=} 0
\end{gather}
yields zeros.
The zeros are obtained in \cite{Sakamoto:2020pev} as
\begin{align}
(y_1, y_2) &= \left(\frac{1/2 + \alpha_2}{M}, \hspace{1pt} \frac12 - \frac{j + \alpha_1}{M} \right), \left(\frac{3/2 + \alpha_2}{M}, \hspace{1pt} \frac12 - \frac{j + \alpha_1}{M} \right), \notag \\
& \hspace{140pt} ..., \left(\frac{(2M - 1)/2 + \alpha_2}{M}, \hspace{1pt} \frac12 - \frac{j + \alpha_1}{M} \right).
\label{zeros}
\end{align}
An important observation here is that the label $j$ characterizes the positions of the zeros and vice versa.
Because one can easily see that the order of zero is always one for all $M$, $j = 0, 1, ..., M-1$ and $(\alpha_1, \alpha_2)$, one finds
\begin{gather}
\frac1{2 \pi i} \oint_{C_i} \bm{\nabla} (\log \, \xi^{j}(z, \tau)) \cdot d\bm{\ell} = + 1 \qquad (i=1, 2, ..., M) \\
\Rightarrow \, \sum_{i=1}^M \chi_i = \sum_{i=1}^M \frac1{2 \pi i} \oint_{C_i} \bm{\nabla} (\log \, \xi^{j}(z, \tau)) \cdot d\bm{\ell} = + M.
\label{index=M}
\end{gather}
Hence, the winding number defined in \eqref{contour_integral} leads to a consistent result $n_+=M$ ($> 0$) with the index theorem \eqref{ASindextheorem}.

Another way to prove \eqref{contour_integral} is to choose a larger contour $C$ containing all the contours $C_i$ and evaluate its contour integral.
Along a parallelogram contour $C\hspace{-3.5pt}: z = 0 \to 1 \to 1 + \tau \to \tau \to 0$, one can compute
\begin{align}
{\rm Ind}\,(i \slashed{D}) &= \frac1{2\pi i} \oint_C \bm{\nabla} (\log \, \xi^j(z, \tau) ) \cdot d\bm{\ell} \notag \\
&= \frac1{2 \pi i} \bigg\{ \int_0^1 dy_1 
\left( \frac{1}{\xi^j(y_1, \tau)}\frac{\partial  \xi^j(y_1, \tau)}{\partial y_1}  
- \frac{1}{\xi^j(y_1 + \tau, \tau)}\frac{\partial  \xi^j(y_1 + \tau, \tau)}{\partial y_1} \right) \notag \\
& \hspace{45pt} + \int_0^1 dy_2 
\left( \frac{1}{\xi^j(1 + \tau y_2, \tau)}\frac{\partial \xi^j(1 + \tau y_2, \tau)}{\partial y_2}  
- \frac{1}{\xi^j(\tau y_2, \tau)}\frac{\partial \xi^j(\tau y_2, \tau)}{\partial y_2} \right) \bigg\} \notag \\
&= +M,
\label{totalwindingnumber}
\end{align}
with the periodicity conditions \eqref{BC_torus} and their derivatives
\begin{align}
\frac{\partial \xi^j(z + 1, \tau)}{\partial y_2} 
&= e^{iq \Lambda_1(z) + 2\pi i\alpha_{1}} 
\left( \frac{iqf}2 + \frac\partial{\partial y_2} \right) \xi^j(z, \tau), \\
\frac{\partial \xi^j(z + \tau, \tau)}{\partial y_1} 
&= e^{iq \Lambda_2(z) + 2\pi i\alpha_{2}} 
\left( -\frac{iqf}2 + \frac\partial{\partial y_1} \right) \xi^j(z, \tau).
\end{align}
As understood from this computation, the index, namely the total winding numbers $\sum_i \chi_i$ is determined only by the conditions \eqref{BC_torus}.

For later convenience, a group of the zeros shown in \eqref{zeros} for a given $j$ is called {\em original zeros} (up to mod 1)
\begin{gather}
Z^j \equiv \left\{\left(\frac{(2k+1)/2 + \alpha_2}{M}, \hspace{1pt} \frac12 - \frac{j + \alpha_1}{M} \right) \Bigg|\, k \in \mathbb{Z}_M \right\},
\end{gather}
where $[x] = \mathrm{max}\,\{n \in \mathbb{Z} \| n \leq x \}$ denotes the floor function.

\section{Modular transformation and zeros}
\subsection{Modular group}
The modular group is isomorphic to $\mathrm{SL}(2, \mathbb{Z})$ and defined as
\begin{gather}
\Gamma = \left\{
\begin{pmatrix}
a & b \\
c & d
\end{pmatrix}
\Bigg|\, a, b, c, d \in \mathbb{Z}, \hspace{1pt} ad -bc = 1 \right\}.
\end{gather}
Indeed, the condition $ad-bc=1$ preserves the volume of the torus $T^2$.
A general transformation acts on the complex modulus $\tau$ and the complex coordinate $z$ as
\begin{gather}
\tau \xmapsto{~\gamma~} \frac{a \tau + b}{c \tau + d} \equiv \gamma \tau, \qquad z \xmapsto{~\gamma~} \frac{z}{c \tau + d} \equiv \gamma z, \\
\gamma =
\begin{pmatrix}
a & b \\
c & d
\end{pmatrix}
\label{gamma}
\in \Gamma,
\end{gather}
where $\tau$ ($z$) and $\gamma \tau$ ($\gamma z$) are regarded as equivalent, respectively.
For convenience, generators of this group are typically chosen as
\begin{gather}
S = 
\begin{pmatrix}
0 & 1 \\ -1 & 0
\end{pmatrix}, \qquad 
T = \begin{pmatrix}
1 & 1 \\ 0 & 1
\end{pmatrix}
\end{gather}
such that 
\begin{gather}
S^4 = (ST)^3 = \mathbbm{1}.
\end{gather}
Consequently, the modular group is generated by the two generators
\begin{gather}
\Gamma = \{ S,T \| S^4 = (ST)^3 = \mathbbm{1} \},
\end{gather}
and the generators $S$ and $T$ act as
\begin{align}
(z, \tau) &\xmapsto{~T~} (z, \tau + 1), \label{trafo_T}\\
(z, \tau) &\xmapsto{~S~} \left( -\frac{z}{\tau}, -\frac1{\tau} \right).
\label{trafo_S}
\end{align}

In this paper, we mainly adopt a definition of the modular group $\Gamma$ as $\mathrm{SL}(2,\mathbb{Z})$. We mention another definition, $\Gamma' = \mathrm{PSL}(2,\mathbb{Z}) = \mathrm{SL}(2,\mathbb{Z})/\{\pm \mathbbm{1}\}$ and discuss its implications in Subsec.\,\ref{PSL}.

\subsection{On periodicity conditions}
In this subsection, we revisit the $S$ and $T$ transformations for the zero-mode wave functions \eqref{sol_torus}.
One originality in this paper is to correctly take the non-zero SS twists $\alpha_i \,\, (i=1,2)$ into account.
That is inspired by an observation that the twists influence the number of independent (zero-mode) physical states on the orbifolds \cite{Abe:2013bca}.

So far in the literature, there exists a point of controversy about the (pseudo-)periodicity conditions \eqref{BC_torus} under lattice translations: whether $S$- and $T$-transformed wave functions should satisfy the original ones or not.
The past conclusions in the literature are as follows,
\begin{itemize}
\item \cite{Ohki:2020bpo, Kikuchi:2020frp}: the boundary conditions \eqref{BC_torus} are satisfied by $S$-transformed wave functions for all $M$, but not by $T$-transformed ones for odd flux quanta ($M=$ odd),
\item \cite{Almumin:2021fbk}: the conditions \eqref{BC_torus} are valid for both $S$- and $T$-transformed wave functions for all $M$.
\end{itemize}

We start with the $T$ transformation, which acts on the zero modes \eqref{sol_torus} as
\begin{align}
\xi^j(z, \tau \| \alpha_1, \alpha_2) \xmapsto{~T~} \xi_T^{j}(z, \tau \| \alpha^{T}_1, \alpha^{T}_2) &\equiv \xi^j(z, \tau+1 \| \alpha^{T}_1, \alpha^{T}_2) \notag\\
&= \mathcal{N}_T\,e^{i \pi M z \, \Im z / \Im \tau} \, \vartheta
\begin{bmatrix}
\tfrac{j + \alpha^{T}_1}M \\[3pt] -\alpha^{T}_2
\end{bmatrix}
(Mz, M(\tau+1)),
\label{T_wf}
\end{align}
where $\mathcal{N}_T = \mathcal{N}$.
Here, we suppose that the twists can transform under $T$, $\alpha_i \xmapsto{~T~} \alpha^T_i \,\, (i=1,2)$ as $\xi^j \xmapsto{~T~} \xi_T^j$.
We list several properties of Jacobi theta-functions in Appx.\,\ref{formulae} which are used in the present section.
Utilizing \eqref{forT_1}\,--\,\eqref{forT_3} several times for $a = (j + \alpha^{T}_1)/M$, $b = - \alpha^{T}_2$, $c = Mz$ and $d = M\tau$, one obtains
\begin{align}
\xi_T^{j}(z+1, \tau \| \alpha^{T}_1, \alpha^{T}_2) &= e^{i \pi M y_2} e^{2\pi i \alpha^{T}_1} \xi_T^{j}(z, \tau \| \alpha^{T}_1, \alpha^{T}_2), \\
\xi_T^{j}(z+\tau, \tau \| \alpha^{T}_1, \alpha^{T}_2) &= e^{-i \pi M y_1} e^{2\pi i ( \alpha^{T}_2 - \alpha^{T}_1 + \rho/2 )} \xi_T^{j}(z, \tau \| \alpha^{T}_1, \alpha^{T}_2),
\end{align}
with $\rho = M$ (mod 2).
One therefore finds that for a particular choice of the $T$-transformed twists
\begin{gather}
\begin{cases}
\alpha^T_1 = \alpha_1\\
\alpha^T_2 = \alpha_1 + \alpha_2 + \rho/2
\end{cases} \hspace{-10pt},
\label{T_twists}
\end{gather}
the boundary conditions \eqref{BC_torus} remain valid for the $T$-transformed wave functions $\xi^j_T$, and thus the result in \cite{Almumin:2021fbk} is verified.
As seen later, a variety of evidence related to zeros of $\xi_T^j$ and the mode counting formula on the orbifolds supports this choice.
One should notice that a contribution from $\rho/2$ must be absorbed into the new twists even though $\alpha_1 = \alpha_2 = 0$, otherwise the original boundary conditions are spoiled.

Next, the $S$ transformation acts as
\begin{align}
\xi^j(z, \tau \| \alpha_1, \alpha_2) \xmapsto{~S~} \xi_S^{j}(z, \tau \| \alpha^{S}_1, \alpha^{S}_2) &\equiv \xi^j(-z/\tau, -1/\tau \| \alpha^{S}_1, \alpha^{S}_2) \notag\\
&= \mathcal{N}_S \,e^{-i \pi M z \, \Re z / \tau} \, \vartheta
\begin{bmatrix}
\tfrac{j + \alpha^{S}_1}M \\[3pt] -\alpha^{S}_2
\end{bmatrix}
(-Mz / \tau, - M / \tau),
\label{S_wf}
\end{align}
where $\mathcal{N}_S = |\tau|^{-1/2} \mathcal{N}$.
Using \eqref{forT_4} for $a = (j + \alpha^{S}_1)/M$, $b = -\alpha^{S}_2$, $c = -z$, $d = \tau/M$ as well as \eqref{forT_2}\,--\,\eqref{forT_3} repeatedly, one has
\begin{align}
\xi_S^{j}(z+1, \tau \| \alpha^{S}_1, \alpha^{S}_2) &= e^{i \pi M y_2} e^{2\pi i \alpha^{S}_2}  \xi_S^{j}(z, \tau \| \alpha^{S}_1, \alpha^{S}_2), \\
\xi_S^{j}(z+\tau, \tau \| \alpha^{S}_1, \alpha^{S}_2) &= e^{-i \pi M y_1} e^{-2\pi i \alpha^{S}_1}  \xi_S^{j}(z, \tau \| \alpha^{S}_1, \alpha^{S}_2).
\end{align}
Thus, for a particular choice
\begin{gather}
\begin{cases}
\alpha^S_1 = -\alpha_2\\
\alpha^S_2 = \alpha_1
\end{cases} \hspace{-10pt},
\label{S_twists}
\end{gather}
the new mode functions $\xi^j_S$ therefore satisfy the original boundary conditions \eqref{BC_torus} again.
This is the result for $\alpha_1 = \alpha_2 = 0$ in the literature \cite{Ohki:2020bpo, Kikuchi:2020frp, Almumin:2021fbk}.
Note that by construction, the same holds for any $\alpha_1$ and $\alpha_2$.
It shall be seen that the choice \eqref{S_twists} is also supported by several evidences.


\subsection{Zeros}
To discuss necessary conditions on zeros in the next sections, we now analyze the zeros of the modular-transformed mode functions $\xi^j_T$ and $\xi^j_S$.

Analogously to Subsec.\,\ref{zero_torus}, by focusing on the Jacobi theta-function appearing in \eqref{T_wf} and solving the equation
\begin{gather}
\vartheta
\begin{bmatrix}
\tfrac{j + \alpha^{T}_1}M \\[3pt] -\alpha^{T}_2
\end{bmatrix}
(Mz, M(\tau+1)) \stackrel{!}{=} 0,
\end{gather}
one finds the zeros of $\xi^j_T$ as
\begin{align}
&(y_1, y_2) = \notag\\
&\left(\frac12 + \frac{1/2-j-\alpha^{T}_1+\alpha^{T}_2}M, \hspace{1pt} \frac12 - \frac{j + \alpha^{T}_1}{M} \right), 
\left(\frac12 + \frac{3/2-j-\alpha^{T}_1+\alpha^{T}_2}M , \hspace{1pt} \frac12 - \frac{j + \alpha^{T}_1}{M}\right), \notag\\
&\hspace{110pt} ..., \left(\frac12 + \frac{(2M-1)/2-j-\alpha^{T}_1+\alpha^{T}_2}M , \hspace{1pt} \frac12 - \frac{j + \alpha^{T}_1}{M}\right).
\end{align}
Solving the equation in the same way
\begin{gather}
\vartheta
\begin{bmatrix}
\tfrac{j + \alpha^{S}_1}M \\[3pt] -\alpha^{S}_2
\end{bmatrix}
(-Mz/\tau, -M/\tau) \stackrel{!}{=} 0,
\end{gather}
one obtains the zeros of $\xi^j_S$ as
\begin{gather}
(y_1, y_2) = \left(\frac12 - \frac{j + \alpha^{S}_1}{M}, \hspace{1pt} \frac{1/2-\alpha^{S}_2}M \right), \left(\frac12 - \frac{j + \alpha^{S}_1}{M}, \hspace{1pt} \frac{3/2-\alpha^{S}_2}M \right), \notag\\
\hspace{140pt} ..., \left(\frac12 - \frac{j + \alpha^{S}_1}{M}, \hspace{1pt} \frac{(2M-1)/2-\alpha^{S}_2}M \right).
\end{gather}

An important implication is that the $S$ transformation does not preserve a set of zeros, even though we adopt the replacement \eqref{S_twists}.
Namely, for the zeros of $\xi^j_T$ and $\xi^j_S$ (up to mod 1)
\begin{align}
Z^j_T &\equiv \left\{ \left(\frac12 + \frac{(2k+1)/2 - j-\alpha^{T}_1+\alpha^{T}_2}M , \hspace{1pt} \frac12 - \frac{j + \alpha^{T}_1}{M} \right) \Bigg|\, k \in \mathbb{Z}_M \right\}, \\
Z^j_S &\equiv \left\{ \left(\frac12 - \frac{j + \alpha^{S}_1}{M}, \hspace{1pt} \frac{(2k+1)/2 - \alpha^{S}_2}M \right) \Bigg|\, k \in \mathbb{Z}_M \right\},
\label{zero_S}
\end{align}
it implies 
\begin{gather}
Z^j = Z^j_T \neq Z^j_S.
\label{zero_unpreserved}
\end{gather}
Note that this consequence does not contradict the index theorem \eqref{contour_integral}.
That is because the index is computed only by the boundary conditions \eqref{BC_torus}, which are valid for $\xi^j_T$ and $\xi^j_S$ as long as we impose \eqref{T_twists} and \eqref{S_twists}.

\section{Modular symmetry}
\subsection{Equivalence of zeros}
It was explained in Sec.\,3 that the generators $S$ and $T$ satisfy $S^4 = (ST)^3 = \mathbbm{1}$.
This can be rephrased as a necessity that the transformed wave functions by $S^4$ and $(ST)^3$ must be identical to the original ones on $T^2$.
We therefore turn to such wave functions and their zeros.

Under an identical transformation $S^4$, the mode functions transform as 
\begin{align}
\xi^j(z, \tau \| \alpha_1, \alpha_2) \xmapsto{~S^4~} \xi_{S^4}^{j}(z, \tau \| \alpha^{S^4}_1, \alpha^{S^4}_2) &= \xi^j(z, \tau \| \alpha^{S^4}_1, \alpha^{S^4}_2) \notag\\
&= \mathcal{N} \,e^{i \pi M z \, \Im z / \Im \tau} \, \vartheta
\begin{bmatrix}
\tfrac{j + \alpha^{S^4}_1}M \\[3pt] -\alpha^{S^4}_2
\end{bmatrix}
(Mz, M\tau).
\end{align}
The zeros then read
\begin{gather}
Z^j_{S^4} \equiv \left\{ \left(\frac{(2k+1)/2 + \alpha^{S^4}_2}M, \hspace{1pt} \frac12 - \frac{j + \alpha^{S^4}_1}M \right) \Bigg|\, k \in \mathbb{Z}_M \right\}
\end{gather}
up to mod 1. Analogously, under another identical transformation $(ST)^3$, one has
\begin{align}
\xi^j(z, \tau \| \alpha_1, \alpha_2) \xmapsto{~(ST)^3~} \xi_{(ST)^3}^{j}(z, \tau \| \alpha^{(ST)^3}_1, \alpha^{(ST)^3}_2) &= \xi^j(z, \tau \| \alpha^{(ST)^3}_1, \alpha^{(ST)^3}_2) \notag\\
&= \mathcal{N} \,e^{i \pi M z \, \Im z / \Im \tau} \, \vartheta
\begin{bmatrix}
\tfrac{j + \alpha^{(ST)^3}_1}M \\[3pt] -\alpha^{(ST)^3}_2
\end{bmatrix}
(Mz, M\tau),
\end{align}
and
\begin{gather}
Z^j_{(ST)^3} \equiv \left\{ \left(\frac{(2k+1)/2 + \alpha^{(ST)^3}_2}M, \hspace{1pt} \frac12 - \frac{j + \alpha^{(ST)^3}_1}M \right) \Bigg|\, k \in \mathbb{Z}_M \right\}.
\end{gather}

Recalling the discussion in Sec.\,4, we know that a combination of the flux quanta $M$, the label $j$, the twists $\alpha_1, \alpha_2$ determines the positions of the zeros and vice versa.
Hence, the original zeros of $\xi^j$ have to be kept equivalent to those of $\xi^j_{S^4}$ and $\xi^j_{(ST)^3}$, i.e.,
\begin{gather}
Z^j = Z^j_{S^4} = Z^j_{(ST)^3} \qquad \forall \, j \in \mathbb{Z}_M.
\end{gather}
This is the necessity of zero-point equivalence, as mentioned previously, and also one evidence that the modular-transformed wave functions satisfy the same boundary conditions as the original ones \eqref{BC_torus}.
This necessity can be rephrased as a restriction on the twists,
\begin{gather}
\begin{cases}
\alpha^{S^4}_1 = \alpha^{(ST)^3}_1 =\alpha_1 \\
\alpha^{S^4}_2 = \alpha^{(ST)^3}_2 =\alpha_2
\end{cases} \hspace{-10pt}.
\end{gather}
One can show that this restriction on the twists under the identical transformations $S^4 = (ST)^3 = \mathbbm{1}$ keeps a consistency with \eqref{T_twists} and  \eqref{S_twists},
\begin{align}
(\alpha_1, \alpha_2) &\xmapsto{~S~} (-\alpha_2, \alpha_1) \notag\\
&\xmapsto{~S~} (-\alpha_1, -\alpha_2) \notag\\
&\xmapsto{~S~} (\alpha_2, -\alpha_1) \notag\\
&\xmapsto{~S~} (\alpha_1, \alpha_2) = (\alpha^{S^4}_1, \alpha^{S^4}_2),
\label{5.7}
\end{align}
and
\begin{align}
(\alpha_1, \alpha_2) &\xmapsto{~T~} (\alpha_1, \alpha_1+\alpha_2+\rho/2) \notag\\
&\xmapsto{~S~} (-\alpha_2,\alpha_1 -\alpha_2 + \rho/2) \notag\\
&\xmapsto{~T~} (-\alpha_1 - \alpha_2 - \rho/2, -\alpha_2) \notag\\
&\xmapsto{~S~} (-\alpha_1 + \alpha_2 - \rho/2, -\alpha_1) \notag\\
&\xmapsto{~T~} (\alpha_2, -\alpha_1) \notag\\
&\xmapsto{~S~} (\alpha_1, \alpha_2) = (\alpha^{(ST)^3}_1, \alpha^{(ST)^3}_2).
\end{align}
Obviously, these relations hold for arbitrary $\alpha_1, \alpha_2 \in [0, 1)$.

Note that the replacements \eqref{T_twists} and \eqref{S_twists} are not necessary at this stage.
Namely, there is no necessity that the common periodicity conditions have to be valid even after the modular transformation, at least on the torus.
However, this turns out to be necessary on the orbifolds, as we shall see in the next section.

\subsection{Comments on $\mathrm{PSL}(2,\mathbb{Z})$}
\label{PSL}
Before closing this section, we mention another definition of the modular group.
See for example \cite{Liu:2019khw}. 
Instead of $\mathrm{SL}(2,\mathbb{Z})$, some authors define the modular group as the projective $\mathrm{SL}(2,\mathbb{Z})$, i.e., $\mathrm{PSL}(2,\mathbb{Z})$
\begin{gather}
\Gamma' = \Gamma/\{\pm \mathbbm{1}\} = \{S, T \| S^2 = -\mathbbm{1}, \hspace{1pt} (ST)^3 = \mathbbm{1}\}/\{\pm \mathbbm{1}\},
\end{gather}
where $\gamma \in \Gamma$ is identified with $-\gamma$.

Unfortunately, our approach in terms of zeros does not distinguish $\mathrm{PSL}(2,\mathbb{Z})$ from $\mathrm{SL}(2,\mathbb{Z})$.
Let us consider a transformation $S^2$ for the zero-mode wave functions,
\begin{align}
\xi^j(z, \tau \| \alpha_1, \alpha_2) \xmapsto{~S^2~} \xi_{S^2}^{j}(z, \tau \| \alpha^{S^2}_1, \alpha^{S^2}_2) &= \xi^j(-z, \tau \| \alpha^{S^2}_1, \alpha^{S^2}_2) \notag\\
&= \mathcal{N} \,e^{i \pi M z \, \Im z / \Im \tau} \, \vartheta
\begin{bmatrix}
\tfrac{j + \alpha^{S^2}_1}M \\[3pt] -\alpha^{S^2}_2
\end{bmatrix}
(-Mz, M\tau).
\end{align}
Then, the zeros are obtained as
\begin{gather}
Z^j_{S^2} \equiv \left\{ \left(\frac{(2k+1)/2 + \alpha_2}M, \hspace{1pt} \frac12 - \frac{-j + \alpha_1}M \right) \Bigg|\, k \in \mathbb{Z}_M \right\},
\end{gather}
where $\alpha^{S^2}_i = -\alpha_i \,\, (i=1,2)$ has been applied (see \eqref{5.7}).
It is now obvious that this set is not identical to the original one $Z^j$.
However, one finds
\begin{gather}
\bigcup_{j \in \mathbb{Z}_M} Z^j = \bigcup_{j \in \mathbb{Z}_M} Z^j_{S^2} = \bigcup_{j \in \mathbb{Z}_M} Z^j_{(ST)^3}.
\end{gather}
Thus, the zeros are invariant under another version of the modular transformation as long as all family indices $j \in \mathbb{Z}_M$ are treated equally.
For instance, nothing gives especial weight to $j$ in the framework of the 6d abelian gauge theory, and then the theory is modular symmetric as a whole.
In contrast, it is well-known that the 10d super Yang-Mills theory appearing as an effective field theory of D9-branes contains non-zero Yukawa interactions among matter fields.
Those may distinguish the labels $j$, since the spinor fields with different labels couple to other matter fields differently.
It depends on concrete setups of flux configurations, and one should be careful of this issue.

\section{Implications on $T^2/\mathbb{Z}_N$}
\subsection{Winding numbers on $T^2/\mathbb{Z}_N$}
We move to the orbifolds $T^2/\mathbb{Z}_N$.
We first summarize the results in the previous paper \cite{Sakamoto:2020pev}.
There, the authors have discovered a zero-mode counting formula.
It counts the number of independent orbifold zero modes on all the orbifolds $T^/\mathbb{Z}_N$ for arbitrary pattern of the flux quanta $M$, the Scherk-Schwarz twists $\alpha_1, \alpha_2$, and the $\mathbb{Z}_{N}$ eigenvalue $\eta$.

In this section, we treat the fixed points \eqref{fixedpoint} in the complex coordinate $z$,
\begin{align}
p_1 = 0, \quad p_2 = 1/2, \quad p_3 = \tau/2, \quad p_4 = (1+\tau)/2 &\qquad \text{on~} T^{2}/\mathbb{Z}_{2}, \label{Z_2fp}\\
p_1 = 0, \quad p_2 = (2+\tau)/3, \quad p_3 = (1+2\tau)/2 &\qquad \text{on~} T^{2}/\mathbb{Z}_{3}, \\
p_1 = 0, \quad p_2 = (1+i)/2, \quad p_3 = 1/2, \quad p_4 = i/2 &\qquad \text{on~} T^{2}/\mathbb{Z}_{4}, \\
\left\{ \hspace{-3pt} 
\begin{array}{c}
p_1 = 0, \quad p_2 = (1+\tau)/3, \quad p_3 = 2(1+\tau)/3, \\[3pt]
p_4 = 1/2, \quad p_5 = \tau/2, \quad p_6 = (1+\tau)/2 
\end{array}\right.
&\qquad \text{on~} T^{2}/\mathbb{Z}_{6}.
\label{Z_6fp}
\end{align}
Here, in addition to the typical fixed points, we have taken into account ``$\mathbb{Z}_2$ fixed points" on $T^2/\mathbb{Z}_4$, and also ``$\mathbb{Z}_3 $ and $\mathbb{Z}_2$ fixed points" on $T^2/\mathbb{Z}_6$, which are not invariant under the $\mathbb{Z}_N$ rotation, but invariant under its partial transformations $\mathbb{Z}_2$ and $\mathbb{Z}_3$.
Such partial fixed points play an important role in the counting formula on the orbifolds.
This is because there exist localized fluxes at some orbifold singularities.
It is investigated in \cite{Buchmuller:2015eya, Buchmuller:2018lkz} that the mode functions vanish at the orbifold singularities where flux is localized.
We will ignore zeros on the bulk in what follows.

Next, let us consider the orbifold eigen functions \eqref{sol_orbifold} and drop unnecessary arguments $\tau$, $\alpha_1$ and $\alpha_2$ for simplicity.
The eigen functions are classified by the $\mathbb{Z}_N$ eigenvalue as
\begin{gather}
\xi^j_\eta (\omega z) = \eta \, \xi^j_\eta (z)
\label{BC_orbifold2}
\end{gather}
with $\eta = \omega^k \,\, (k=0,1,...,N-1)$ and $\omega = e^{2 \pi i/N}$.
Given the (pseudo-)periodicity conditions \eqref{BC_torus} and the reflection \eqref{BC_orbifold2}, one can easily obtain for a given fixed point $p_i$ the form
\begin{gather}
\xi^j_\eta (\omega z + p_i) = e^{2 \pi i w/N} \, \xi^j_\eta(z + p_i), \qquad \omega = e^{2 \pi i/N}.
\label{6.6}
\end{gather}
Then, the winding number $\chi_i$ around $z = p_i$ is precisely equal to $w$ (mod $N$) \cite{Sakamoto:2020pev}.
For ``$\mathbb{Z}_{N'}$ fixed points", we exceptionally adopt $N'$ instead of $N$.
We list the winding numbers on each orbifold,
\begin{align}
\left\{ \hspace{-3pt} 
\begin{array}{c}
\chi_1 = k, \quad \chi_2 = k - 2\alpha_1, \quad \chi_3 = k -2 \alpha_2, \\[3pt]
\chi_4 = k - 2(\alpha_1 + \alpha_2) -M \quad (\text{mod~} 2)
\end{array}\right.
&\qquad \text{on~} T^{2}/\mathbb{Z}_{2}, \label{winding_N=2}\\
\chi_1 = k, \quad \chi_2 = -M - 6\alpha + k, \quad \chi_3 = 2M + 6\alpha + k \quad (\text{mod~} 3)
&\qquad \text{on~} T^{2}/\mathbb{Z}_{3}, \\
\left\{ \hspace{-3pt} 
\begin{array}{c}
\chi_1 = k, \quad \chi_2 = M - 4\alpha + k \quad (\text{mod~} 4), \\[3pt]
\chi_3 = \chi_4 = - 2\alpha + k \quad (\text{mod~} 2)
\end{array}\right.
&\qquad \text{on~} T^{2}/\mathbb{Z}_{4}, \\
\left\{ \hspace{-3pt} 
\begin{array}{c}
\chi_1 = k \quad (\text{mod~} 6), \\[3pt]
\chi_2 = \chi_3 = M/2 - 3\alpha -2k \quad (\text{mod~} 3),\\[3pt]
\chi_4 = \chi_5 = \chi_6 = - 2\alpha - k \quad (\text{mod~} 2)
\end{array}\right.
&\qquad \text{on~} T^{2}/\mathbb{Z}_{6}.
\label{winding_N=6}
\end{align}
The explicit expressions for \eqref{6.6} are thoroughly listed in Appx.\,\ref{formula_winding}.

\subsection{Invariance of zero-mode counting formula}
We turn to the zero-mode counting formula on the magnetized orbifolds $T^2/\mathbb{Z}_N$.
In terms of the winding numbers obtained previously, the number of independent orbifold zero modes, $n$, is counted by \cite{Sakamoto:2020pev}
\begin{gather}
n = \frac{M-V}N + 1,
\label{countingformula}
\end{gather}
where $V \equiv \sum_i \chi_i$ denotes the sum of winding numbers around the orbifold singularities.
Note that $V$ can be calculated only by the translation conditions \eqref{BC_torus} and the reflection one \eqref{BC_orbifold2}, by construction.
Hence, as long as we adopt \eqref{T_twists} and \eqref{S_twists}, the counting formula is symmetric under the modular transformations $S$ and $T$ in $\Gamma$ (as well as $\Gamma'$).

We finally give a comment on a physical meaning of the orbifold fixed points, geometrically and gauge-theoretically.
As is well-known, the fixed points have a geometrical meaning, since infinite curvatures exist at them (as well as the partial fixed points).
Furthermore, there are localized flux sources at them from the gauge-theoretical point of view \cite{Buchmuller:2015eya, Buchmuller:2018lkz}.
Hence, the fixed points listed in \eqref{Z_2fp}\,--\,\eqref{Z_6fp} do have a physical meaning.
Recall that the fixed points $z_\mathrm{fp}$ are the solution of an equation
\begin{gather}
z_\mathrm{fp} = \omega z_\mathrm{fp} + m + n \tau
\end{gather}
for $\exists \, m, n \in \mathbb{Z}$.
Here, the both sides transform under the modular transformation, and the fixed points \eqref{Z_2fp}\,--\,\eqref{Z_6fp} remain as fixed points again.
Although its connection to the index theorem has not been clarified in the literature, {\em a priori} the zero-mode counting formula \eqref{countingformula} is modular-invariant.
Notice that it is not invariant if we exclude \eqref{T_twists} and \eqref{S_twists}.
For instance, one then finds an incorrect result after some algebra: the points $p_2$ and $p_3$ do not behave as fixed points, and $\chi_2 = \chi_3 = 0$ for $N=3$, $M=6m+2 \,\, (m \in \mathbb{N})$, $\eta = \omega^2$ and $\alpha=1/3$.
Hence, the modular symmetry of the counting formula necessitates that the twist phases $\alpha_i \,\, (i=1,2)$ appropriately transform as \eqref{T_twists} and \eqref{S_twists} under the $T$ and $S$ transformations.

This is subtle, but would give a correct understanding for the modular symmetry and shed new light on modular flavor symmetries.

\section{Conclusion}
In this paper, we have considered the two-dimensional torus $T^2$ and the toroidal orbifolds $T^2/\mathbb{Z} \,\, (N=2,3,4,6)$ with a background magnetic field.
Inspired by the recent papers \cite{Ohki:2020bpo, Kikuchi:2020frp, Almumin:2021fbk}, we have revisited the modular group in magnetic compactifications.
We have focused on the zeros of the ground-state mode functions and looked at the transformation property of the zeros.
We have also examined what type of boundary conditions must be satisfied after the modular transformation.
A crucial observation is that
\begin{itemize}
\item the Scherk-Schwarz twist phases $\alpha_i \,\, (i=1,2)$ transform under the $T$ and $S$ transformations if and only if the (pseudo-)periodicity conditions \eqref{BC_torus} are valid even after the modular transformation.
\end{itemize}
This observation precisely agrees with two necessities
\begin{itemize}
\item the zeros are also identical under two identical transformations $S^4 = (ST)^3 = \mathbbm{1}$,
\item the zero-mode counting formula on the orbifolds is modular symmetric.
\end{itemize}
Furthermore, we have mentioned another definition of the modular group as $\mathrm{PSL}(2,\mathbb{Z})$, and it turned out that our approach in terms of zeros yields the same implication as that of $\mathrm{SL}(2, \mathbb{Z})$.

In our conclusion, it is reasonable to adopt the twists \eqref{T_twists} and \eqref{S_twists} after the $T$ and $S$ transformations.
Thus, the conclusion justifies the claim in \cite{Almumin:2021fbk} that the periodicity conditions \eqref{BC_torus} is common under the modular transformation.
A crucial consequence in this paper is that the replacements \eqref{T_twists} and \eqref{S_twists} are justified only after we take the positions of zeros and their modular-equivalence into account.
It would help further analyses on modular flavor symmetries and their phenomenological applications.
We will pursue them somewhere.

\section*{Acknowledgment}
I would like to thank Kantaro Ohmori for helpful comments.
The work is supported in part by Scuola Normale, by INFN (IS GSS-Pi) and by the MIUR-PRIN contract 2017CC72MK\_003.

\appendix
\renewcommand{\thesection}{\Alph{section}}

\section{Gamma matrices}
In this paper, we follow the notation in \cite{Abe:2013bca, Sakamoto:2020pev}.
For $M,N = \mu \, (= 0, 1, 2, 3), 5, 6$, we adopt the 6d metric and the gamma matrices
\begin{gather}
\{ \Gamma^M, \Gamma^N \} = 2 \eta^{MN}, \\
\eta^{MN} = {\rm diag} \, (+1, -1, -1, -1, -1, -1), \\
\Gamma^\mu = 
\begin{pmatrix}
\gamma^\mu & 0 \\
0 & \gamma^\mu
\end{pmatrix}, \\
\Gamma^5 = 
\begin{pmatrix}
0 & i\gamma_5 \\
\gamma_5 & 0
\end{pmatrix}, \qquad 
\Gamma^6 = 
\begin{pmatrix}
0 & \gamma_5 \\
-\gamma_5 & 0
\end{pmatrix}, \qquad 
\Gamma^7 = 
\begin{pmatrix}
\gamma_5 & 0 \\
0 & - \gamma_5
\end{pmatrix}.
\end{gather}

\section{Jacobi theta-functions}
\label{formulae}
We list the properties of Jacobi theta-functions (cf.\,\cite{Choi:2006qh}) which are used in the previous sections.

The Jacobi theta-function with characteristics $a$ and $b$ is defined as
\begin{gather}
\vartheta
\begin{bmatrix}
a \\[3pt] b
\end{bmatrix}
(c, d)
 = \sum_{l=-\infty}^{\infty} e^{\pi i(a+l)^{2}d}\,e^{2\pi i(a+l)(c + b)}.
\end{gather}
It satisfied the relations
\begin{align}
\vartheta
\begin{bmatrix}
a \\[3pt] b
\end{bmatrix}
(c, d + 1) &= e^{-i \pi (a^2 - a)} ~ 
\vartheta
\begin{bmatrix}
a \\[3pt] b + a - \tfrac12
\end{bmatrix}
(c, d),
\label{forT_1}\\
\vartheta
\begin{bmatrix}
a \\[3pt] b
\end{bmatrix}
(c + 1, d) &= e^{2\pi i a} ~ 
\vartheta
\begin{bmatrix}
a \\[3pt] b
\end{bmatrix}
(c, d), 
\label{forT_2}\\
\vartheta
\begin{bmatrix}
a \\[3pt] b
\end{bmatrix}
(c + d, d) &= e^{-2\pi i (b + c + d/2)} ~ 
\vartheta
\begin{bmatrix}
a \\[3pt] b
\end{bmatrix}
(c, d),
\label{forT_3}
\end{align}
and these are used to compute the (pseudo-periodicity) boundary conditions for $\xi^j_T$.
Using an identity
\begin{gather}
\vartheta
\begin{bmatrix}
a \\[3pt] b
\end{bmatrix}
(c/d, -1/d) = \sqrt{-id} \, e^{2 \pi i (c^2/2d + ab)} ~ 
\vartheta
\begin{bmatrix}
b \\[3pt] -a
\end{bmatrix}
(c, d),
\label{forT_4}
\end{gather}
one can reduce some complicated computations to easier ones in terms of \eqref{forT_2} and \eqref{forT_3}.

\section{More on Eq.\,\eqref{6.6}}
\label{formula_winding}
We thoroughly show the explicit expressions for \eqref{6.6} on each orbifold $T^2/\mathbb{Z}_N$, based on \cite{Sakamoto:2020pev}.
For all the orbifolds and $\omega = e^{2\pi i/N} \,\, (N=2,3,4,6)$, the reflection around the origin $z = 0$ determines the $\mathbb{Z}_N$ eigenvalue (or parity) $\eta \equiv \omega^k$,
\begin{gather}
\xi_{\omega^{k}} (\omega z) = \omega^{k} \, \xi_{\omega^{k}} (z) \qquad (k = 0, 1, \ldots, N-1).
\end{gather}
For the other orbifold singularities, one finds for $N=2$
\begin{align}
\xi_{\omega^{k}} (-z + \tfrac12) &= e^{-iq \Lambda_1(z) - 2 \pi i (\alpha_1 -k/2)} \, \xi_{\omega^{k}} (z + \tfrac12), \\
\xi_{\omega^{k}} (-z + \tfrac{\tau}2) &= e^{-iq \Lambda_2(z) - 2\pi i (\alpha_2-k/2)} \, \xi_{\omega^{k}} (z + \tfrac{\tau}2), \\
\xi_{\omega^{k}} (-z + \tfrac12 + \tfrac{\tau}2) &= e^{-iq \Lambda_1(z) - iq \Lambda_2(z) - 2\pi i (M/2 + \alpha_1 + \alpha_2 -k/2)} \, \xi_{\omega^{k}} (z + \tfrac12 + \tfrac{\tau}2).
\end{align}
Furthermore, one can derives for $N=3$ in the same way
\begin{align}
\xi_{\omega^{k}} (\omega z + \tfrac23 + \tfrac{\tau}3) &= e^{- iq \Lambda_1(z) - iq \Lambda_2(z) - 2 \pi i (M/3 + 2\alpha - k/3 )} \, \xi_{\omega^{k}} (z + \tfrac23 + \tfrac{\tau}3 ), \\
\xi_{\omega^{k}} (\omega z + \tfrac13 + \tfrac{2\tau}3) &= e^{iq \Lambda_1(\omega z) + iq \Lambda_2(\omega z) + 2 \pi i (2M/3 + 2\alpha + k/3 )} \, \xi_{\omega^{k}} (z + \tfrac13 + \tfrac{2\tau}3 ),
\end{align}
for $N=4$,
\begin{align}
\xi_{\omega^k} (\omega z + \tfrac12 + \tfrac{\tau}2) &= e^{-iq \Lambda_2(z) - 2 \pi i (-M/4 + \alpha - k/4)} \, \xi_{\omega^k} (z + \tfrac12 + \tfrac{\tau}2), \\
\xi_{\omega^k} (\omega^2 z + \tfrac12) &= e^{-iq \Lambda_1(z) - 2 \pi i (\alpha -k/2 )} \, \xi_{\omega^k} (z + \tfrac12), \\
\xi_{\omega^k} (\omega^2 z + \tfrac{\tau}2) &= e^{-iq \Lambda_2(z) - 2\pi i (\alpha - k/2)} \, \xi_{\omega^k} (z + \tfrac{\tau}2),
\end{align}
and for $N=6$,
\begin{align}
\xi_{\omega^k} (\omega^2 z + \tfrac13 + \tfrac{\tau}3) &= e^{-iq \Lambda_2(z) -2 \pi i (-M/6 + \alpha + 2k/3)} \, \xi_{\omega^k} (z + \tfrac13 + \tfrac{\tau}3), \\
\xi_{\omega^k} (\omega^3 z + \tfrac12) &= e^{-iq \Lambda_1(z) - 2\pi i (\alpha + k/2)} \, \xi_{\omega^k} (z + \tfrac12),
\end{align}
where we have used the relations $\chi_2 = \chi_3$ and $\chi_4 = \chi_5 = \chi_6$ on $T^2/\mathbb{Z}_6$.
Ignoring the factors including $\Lambda_1(z)$ and $\Lambda_2(z)$ for an infinitesimally small $|z|$ yields Eq.\,\eqref{6.6} and the corresponding winding numbers in Eqs.\,\eqref{winding_N=2}\,--\,\eqref{winding_N=6}.

\bibliographystyle{JHEP}
\bibliography{references}
\end{document}